\documentclass[%
 reprint,
 superscriptaddress,
 showpacs,preprintnumbers,
 amsmath,amssymb,
 aps,
 prb,
]{revtex4-1}

\usepackage{graphicx}
\usepackage{dcolumn}
\usepackage{bm}

\usepackage{upgreek}
\usepackage{amsmath}

\begin{document}


\title{Robust enhancement of high-harmonic generation from all-dielectric metasurfaces enabled by polarization-insensitive bound states in the continuum}

\author{Shuyuan Xiao}
\email{syxiao@ncu.edu.cn}
\affiliation{Institute for Advanced Study, Nanchang University, Nanchang 330031, China}
\affiliation{Jiangxi Key Laboratory for Microscale Interdisciplinary Study, Nanchang University, Nanchang 330031, China}

\author{Meibao Qin}
\affiliation{School of Physics and Materials Science, Nanchang University, Nanchang 330031, China}

\author{Junyi Duan}
\affiliation{Institute for Advanced Study, Nanchang University, Nanchang 330031, China}
\affiliation{Jiangxi Key Laboratory for Microscale Interdisciplinary Study, Nanchang University, Nanchang 330031, China}

\author{Tingting Liu}
\email{ttliu@usst.edu.cn}
\affiliation{Institute of Photonic Chips, University of Shanghai for Science and Technology, Shanghai 200093, China}
\affiliation{Centre for Artificial-Intelligence Nanophotonics, School of Optical-Electrical and Computer Engineering, University of Shanghai for Science and Technology, Shanghai 200093, China}

\begin{abstract}
	
The emerging all-dielectric platform exhibits high-quality ($Q$) resonances governed by the physics of bound states in the continuum (BIC) that drives highly efficient nonlinear optical processes. Here we demonstrate the robust enhancement of third-(THG) and fifth-harmonic generation (FHG) from all-dielectric metasurfaces composed of four silicon nanodisks. Through the symmetry breaking, the genuine BIC transforms into the high-$Q$ quasi-BIC resonance with tight field confinement for record high THG efficiency of $3.9\times10^{-4}$ W$^{-2}$ and FHG efficiency of $4.8\times10^{-10}$ W$^{-4}$ using a moderate pump intensity of 1 GW/cm$^{2}$. Moreover, the quasi-BIC and the resonantly enhanced harmonics exhibit polarization-insensitive characteristics due to the special $C_{4}$ arrangement of meta-atoms. Our results suggest the way for smart design of efficient and robust nonlinear nanophotonic devices.

\end{abstract}

\maketitle


\maketitle

Nonlinear frequency conversion, such as harmonic generation that greatly broadens the accessible spectrum, are ubiquitous in applications from our daily life to cutting-edge science and technology\cite{Boyd2020}. Over the past decade, the emergence of artificially structured nanoresonators, i.e., metamaterials and metasurfaces, has revolutionized our perception in nonlinear optics\cite{Li2017}. Recently, all-dielectric metasurfaces provide a paradigm for efficient and flexible harmonic generation in nonlinear nanoscale optics due to the high index, low loss, and resonant field enhancement\cite{Sain2019, Grinblat2021}. For example, the third-harmonic generation (THG) was studied in dielectric metasurfaces made from centrosymmetric semiconductors such as silicon and germanium, and the second-harmonic generation (SHG) was reported in the III–V semiconductor metasurfaces. In particular, the ability to support both electric and magnetic Mie-type resonances is beneficial to enhance nonlinear responses. Various dielectric metasurfaces, supporting the magnetic dipole resonance\cite{Shcherbakov2014, Yang2015, Liu2021}, toroidal dipole resonance\cite{Rocco2018}, and nanoradiating anapole resonance\cite{Grinblat2016, Xu2018, Yao2020}, have been explored to increase the field confinement in tight volumes, resulting in a considerable increase of nonlinear conversion efficiency.

Most recently, a special optical resonant mode originating from the physics of bound states in the continuum (BIC) gains increasing interest in boosting the harmonic generation from dielectric metasurfaces. Optical BIC underpins the existence of perfectly confined states embedded into the radiation spectrum and thus suffers from inaccessibility of external excitations\cite{Hsu2016}. It was reported that dielectric metasurfaces with broken in-plane inversion symmetry can support ultrahigh quality ($Q$) factor resonances arising from distortion of symmetry-protected BICs\cite{Koshelev2018, Li2019, Saadabad2021}. Such quasi-BIC resonances with large field confinement have been employed to significantly enhance nonlinear responses such as THG and SHG by orders of magnitude on dielectric metasurface platform\cite{Vabishchevich2018, Xu2019, Liu2019, Anthur2020, Ning2021}. Especially, efficient high-harmonic generation (HHG) up to 11th order was also experimentally observed from silicon metasurfaces empowered by the symmetry-protected BIC\cite{Zograf2022}. However, a direct consequence of these quasi-BICs excited by introducing structural symmetry perturbation is the sensitivity to polarization angle of the pump beam, which severely hinders the robustness in their practical applicability.

In this letter, we demonstrate the robust enhancement of the harmonic signals from all-dielectric metasurfaces that is not dependent on the incidence polarization. The nonlinear metasurfaces with in-plane symmetry perturbation are designed using the symmetry selection rules such that the quasi-BIC and optical response are polarization independent under normal incidence. The high $Q$ factor resonance with strongly enhanced local fields leads to significant enhancement of the THG and fifth-harmonic generation (FHG) signals, and the nonlinear harmonics show robustness to the polarization angles. Our results have important implications for novel nonlinear optical devices and restore a flexible design strategy for polarization-independent harmonic signals boosted by symmetry-protected BICs. 

\begin{figure}[htbp]
	\centering
	\includegraphics[scale=0.34]{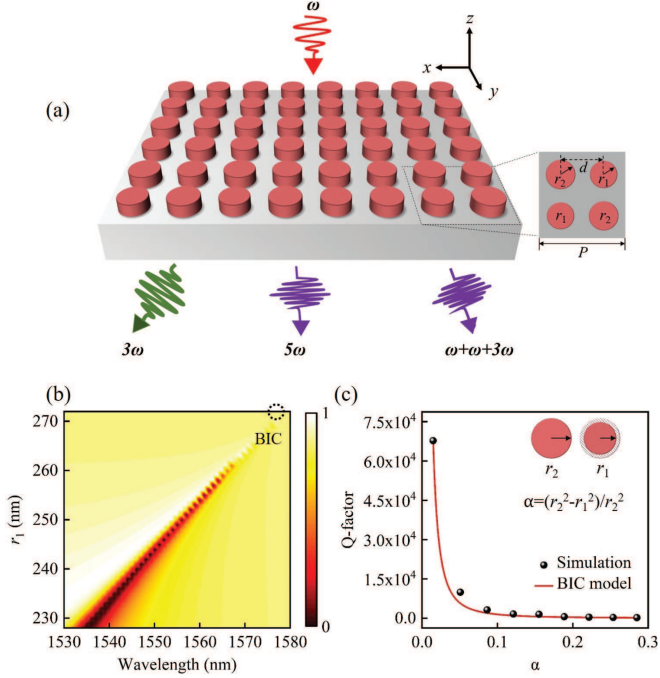}
	\caption{\label{figure1}(a) The schematic illustration of the nonlinear metasurfaces comprising a square lattice of four silicon nanodisks. Photons of the pump beam at frequency $\omega$ are converted to photons at $3\omega$ by THG and $5\omega$ by direct FHG and two-step cascaded FHG. (b) The simulated transmission spectrum of the metasurfaces as a function of pump light wavelength and nanodisk radius $r_1$, under the $x$-polarized normally incident plane waves. The genuine BIC state is marked by a circle when the four nanodisks are identical with $r_1=r_2$. (c) The dependence of $Q$ factor of the quasi-BIC on the asymmetry parameter defined as $\alpha= (r_{2}^2-r_{1}^2)/r_{2}^2$. }
\end{figure}

We consider dielectric metasurfaces comprising complex meta-atoms in the form of four silicon nanodisks on a silica substrate under the normal incident plane wave, as shown in Fig. \ref{figure1}(a). The silicon is adopted for the odd order nonlinear processes including THG, direct FHG, and cascaded FHG. The asymmetric nanodisks are designed to show a twofold rotational symmetry, i.e. $C_{2}$ symmetry. We perform a numerical analysis of the linear response using the finite element method solver in COMSOL Multiphysics in the frequency domain. The metasurfaces are considered as an infinite structure with perfect periodicity. The refractive index of silicon is taken from experiments and that of silica semi-infinite substrate is set as 1.45\cite{Palik1985}. The square lattice is designed with period $P=1400$ nm, nanodisk height $h=200$ nm, center-to-center separation between nanodisks $d=700$ nm, and nanodisk radius $r_1$ varies while $r_2$ is fixed as 272 nm. 

Figure \ref{figure1}(b) shows the transmission spectra of the metasurfaces with nanodisk radius $r_{1}$ ranging from 228 nm to 272 nm under $x$-polarized normally incident plane waves. When their radii are perfectly identical, i.e. $r_{1}=r_{2}$, the metasurfaces support a genuine BIC marked by a circle, since the symmetry is incompatible with modes of free space. As expected, there is no dip in the transmission spectrum for the genuine BIC case, implying a completely decoupling of the mode from external radiation. The energy coupling of such symmetry-protected BIC system necessitates a certain radiation leakage channel, which is achieved through varying $r_{1}$ in this case. In the broken-symmetry metasurfaces, the resonance manifests itself as a narrow dip in the transmission spectrum. In Fig. \ref{figure1}(c), we plot the evolution of the $Q$ factor of quasi-BIC resonance as a function of the asymmetry parameter $\alpha$. Here $\alpha$ is defined by the difference in nanodisk cross sectional area $\Delta S/S$. The radiative $Q$ factor depends on $\alpha$ in the form of the inverse quadratic law, $Q=Q_{0}\alpha^{-2}$, where $Q_{0}$ is a constant determined by the metasurface design independent of $\alpha$\cite{Koshelev2018, Li2019, Saadabad2021}. As it can be seen, the $Q$ factor experiences a dramatic decrease as $\alpha$ increases. Consequently, the nonradiating symmetry-protected BIC with infinite $Q$ factor is effectively transformed into the radiating quasi-BICs with ultrahigh $Q$ factors as the symmetry perturbation is introduced in the metasurfaces.

\begin{figure}[htbp]
	\centering
	\includegraphics[scale=0.40]{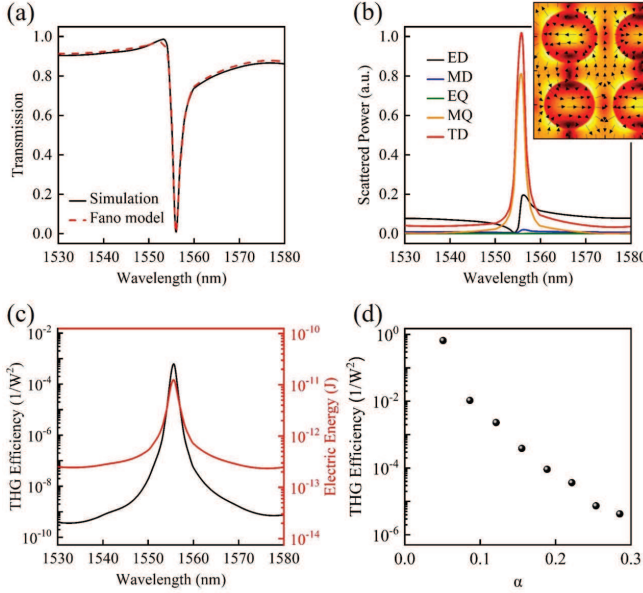}
	\caption{\label{figure2}(a) The transmission spectra of the metasurfaces from simulation and theoretical Fano model. (b) Multipolar decomposition of the linear scattering cross section. The inset shows the near-field distribution with field vectors in the $x$-$y$ plane at the resonance. (c) The normalized conversion efficiency of THG and the electric energy insider silicon nanodisks as a function of the pump wavelength. (d) The THG conversion efficiency as a function of the asymmetry parameter $\alpha$. The $x$-polarized normally incident light is employed in (a)-(d).}
\end{figure}

We use the quasi-BIC resonance to achieve the local field enhancement and highly efficient harmonics. Here the broken-symmetry metasurfaces with $r_{1}$=250 nm ($\alpha=0.15$) are adopted and the quasi-BIC resonant wavelength is designed at 1556 nm. In Fig. \ref{figure2}(a), the transmission spectrum of the metasurfaces demonstrates a quasi-BIC resonance with a Fano asymmetric line shape, which can be fitted by $T=|a_{1}+ia_{2}+\frac{b}{\omega-\omega_{0}+i\gamma}|$, where $a_{1}$, $a_{2}$, and $b$ are constant real numbers, $\omega_{0}$ is the resonant frequency, $\gamma$ is the overall damping rate of the resonance. Then the $Q$ factor can be evaluated by $\omega_{0}/2\gamma$ as 1504. We further perform the multipolar decomposition of the optical linear response to uncover the nature of the quasi-BIC resonance. It is observed in Fig. \ref{figure2}(b) that toroidal dipole (TD) and magnetic quadrupole (MQ) are excited in the metasurfaces, with predominant contributions to quasi-BIC resonance. Such resonant feature can also be clearly observed from the near-field distribution in the inset. Note that the simulated enhancement of the near-field amplitude at resonance can be up to 40 times inside the structures. 

Then we concentrate on the nonlinear responses in the metasurfaces, starting with the THG process from the above-designed configuration. To calculate the harmonic generation from the structure, two coupled electromagnetic models within the undepleted pump approximation are exploited. First, the linear response at pump wavelength is solved for the local field distributions. Second, the induced nonlinear polarization inside the meta-atoms is employed as a source for the electromagnetic simulation at the harmonic wavelength to obtain the harmonic field. For the THG process, the nonlinear polarization in the silicon nnaodisks can be computed by $\bm{P}^{3\omega}=3\varepsilon_{0}\chi^{(3)}(\bm{E\cdot\bm{E}})\bm{E}^{\omega}$, where $\varepsilon_{0}=8.8542\times10^{-12}$ F/m is the vacuum permittivity, $\chi^{(3)}=2.45\times10^{-19}$ m$^{2}$/V$^{2}$ is the third-order susceptibility of silicon, and $\bm{E}^{\omega}$ is the electric field at the pump wavelength. The THG conversion efficiency is normalized as $\eta_{\text{TH}}=P_{\text{TH}}/(P_{\text{FF}})^{3}$, where $P_{\text{TH}}$ is the total radiated power at the third harmonic and $P_{\text{FF}}$ is the pump power incident on the resonator at the fundamental frequency. 

In Fig. \ref{figure2}(c), the THG efficiency reaches its peak of $3.9\times10^{-4}$ W$^{-2}$ at the pump wavelength of 1556 nm corresponding to the position of the quasi-BIC resonance. Compared to wavelengths detuned from the resonance, the peak efficiency at quasi-BIC represents an enhancement of more than six orders of magnitude. Such a record high value of THG efficiency significantly outperforms the previously reported performance of the existing silicon resonators under normal excitation, as listed in Table 1. Essentially, since THG is predominantly a volume phenomenon that responds mainly to the electric component of the incident wave, the electric energy inside the structure at the pump wavelength can be used to interpret the THG efficiency characteristics\cite{Boyd2020}. As reported on the right axis of the Fig. \ref{figure2}(c), the electric energy within the pump wavelength regime exhibits a peak at 1556 nm. This reasonable agreement intuitively illustrates the large enhancement of THG efficiency at the quasi-BIC resonance resulting from the tight energy confinement. To analyze the dependence of the THG efficiency with the symmetry perturbation of the metasurfaces, the normalized efficiencies with respect to the asymmetry parameter $\alpha$ are exhibited in Fig. \ref{figure2}(d). The results show the expected trend of THG efficiency decreasing with the increase of $\alpha$, since large deviation from the symmetric unit cell facilitates the radiation leakage of the symmetry-protected BIC. This behavior reveals the controllable THG efficiency by the magnitude of the symmetry perturbation, and thus the THG performance can be improved by fabricating small defects with specialized process. 

\begin{table}[htbp]
	\centering
	\caption{\label{table1}\bf Comparison of THG performance from silicon resonant structures}
	\begin{tabular}{p{40pt}p{50pt}p{60pt}p{50pt}}
		\hline
		Reference                & Pump wavelength (nm)   & Pump intensity (GW/cm$^2$)   & $\eta_{\text{TH}}$ (W$^{-2}$) \\
		\hline
		\cite{Shcherbakov2014}   & 1260                   & 5                            & $2.6\times10^{-14}$\\
		\cite{Yang2015}          & 1350                   & 3.2                          & $3\times10^{-8}$\\
		\cite{Xu2018}            & 1550                   & 3                            & $3.9\times10^{-7}$\\
		\cite{Carletti2019}      & 1601                   & 1                            & $2.9\times10^{-5}$\\
		\cite{Matsudo2022}       & 1325                   & 0.32                         & $1\times10^{-4}$\\  
		This work                & 1556                   & 1                            & $3.9\times10^{-4}$\\  
		\hline
	\end{tabular}
\end{table}

\begin{figure}[htbp]
	\centering
	\includegraphics[scale=0.40]{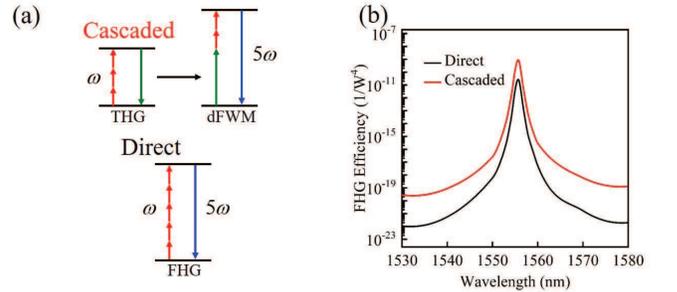}
	\caption{\label{figure3}(a) The schematic of FHG via direct process, or through cascaded THG and dFWM. (b) The normalized conversion efficiency of FHG via direct and cascaded pathway for $x$-polarized normal incidence.}
\end{figure}

The proposed dielectric metasurfaces that host a high $Q$ factor quasi-BIC resonance provides a promising solution to generate high-harmonic signal on the subwavelength scale at moderate driving intensities. Here we consider the potential in boosting FHG process under normally incident plane wave. In a conventional way, a direct process involved with the fifth-order susceptibility $\chi^{(5)}$ is used to obtain the FH signal. Accordingly, this direct FHG process is simulated through the nonlinear polarization $\bm{P}^{5\omega}=3\varepsilon_{0}\chi^{(5)}(\bm{E\cdot\bm{E}})^{2}\bm{E}^{\omega}$, where $\chi^{(5)}$ of silicon is estimated as $5.6\times10^{-39}$ m$^{4}$/V$^{4}$. On the other hand, the FHG can be achieved by a cascaded process of the THG followed by degenerate four wave mixing (dFWM), as illustrated in Fig. \ref{figure3}(a). The cascaded process shows advantages of higher efficiency and extremely short time constant. In the simulation, the polarization of the cascade THG is defined as $\bm{P}^{5\omega}=3\varepsilon_{0}\chi^{(3)}[2(\bm{E^{\omega}\cdot\bm{E}^{3\omega}})\bm{E}^{\omega}+(\bm{E}^{\omega}\cdot\bm{E}^{\omega})\bm{E}^{3\omega}]$, where $\bm{E}^{\omega}$ and $\bm{E}^{3\omega}$ are the electric field at fundamental frequency $\omega$ and TH frequency $3\omega$, respectively. The calculated FHG efficiencies of the two pathways are depicted in Fig. \ref{figure3}(b). At the pump intensity of 1 GW/cm$^{2}$, the cascaded effects show the peak efficiency of $4.8\times10^{-10}$ W$^{-4}$ at the quasi-BIC resonance wavelength, which is higher than that of the direct process of $1.5\times10^{-11}$ W$^{-4}$ due to the relation between $\chi^{(5)}$ and $\chi^{(3)}$ by $\chi^{(5)} \approx [\chi^{(3)}]^2/\chi^{(1)}$. 

\begin{figure}[h!]
	\centering
	\includegraphics[scale=0.40]{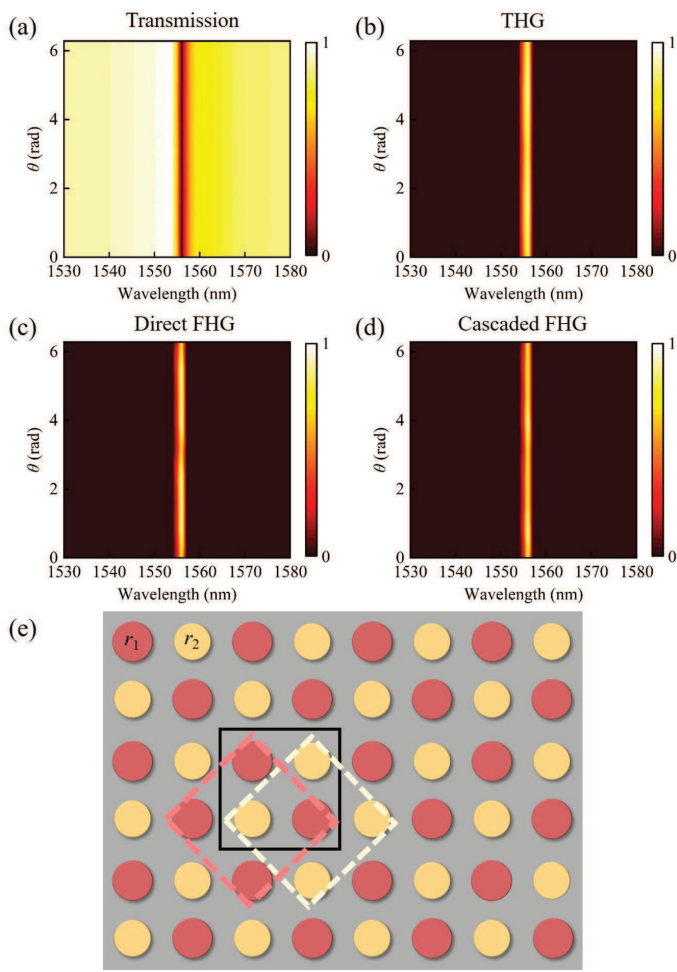}
	\caption{\label{figure4}(a) The transmission spectra, (b) THG efficiencies, (c) direct FHG efficiencies, and (d) cascaded FHG efficiencies of the proposed metasurfaces as a function of the polarization angle varying from 0 to $2\pi$. (e) The schematic of the $C_{2}$ and $C_{4}$ symmetry due to the meta-atom arrangement. The unit cell has $C_{2}$ symmetry (the solid square) and the unit cells have $C_{4}$ symmetry (the dashed square). Nanodisks of the same radii are shown with the same color.}
\end{figure}

As the most intriguing part, we demonstrate that the proposed metasurfaces support polarization-insensitive quasi-BIC resonance and its optical response in either linear or nonlinear regime keeps unchanged under polarization variation. In most previous works that concentrated on BIC-enhanced harmonic generation, the quasi-BIC resonance can only be excited by a specific polarization light and the significant harmonic signals can only be expected in this situation, following the selection rule of the so-called symmetry-protected BIC. Figs. \ref{figure4}(a)-(d) present the linear transmission spectra and the normalized nonlinear harmonic signals of the metasurfaces with various pump polarization angles. It is observed that the transmission spectrum does not change as the polarization angle varies from 0 to $2\pi$, revealing the polarization-insensitive feature of the quasi-BIC here, which leads to the THG and FHG enhancement robust to the polarization angles because of the direct dependence of nonlinear response with linear response. Such robustness can be unveiled by the symmetry in arranging the meta-atoms, as illustrated in Fig. \ref{figure4}(e). As highlighted by the solid square, the unit cell shows $C_{2}$ symmetry. In the meanwhile, the unit cell constructed by neighboring nanodisks with the same radius $r_{1}$ possesses the $C_{4}$ symmetry, and this kind of $C_{4}$ symmetry also applies to the unit cell with nanodisks of radius $r_{2}$, shown by the dashed squares. Since the $C_4$ symmetry is required for the polarization insensitivity, the special arrangement of meta-atoms in the proposed metasurfaces enables the polarization-insensitive optical responses in both the linear and nonlinear regimes. 

In summary, we propose the robust BIC enhanced of THG and FHG in the dielectric metasurfaces composed of four silicon nanodisks. Through the symmetry perturbation in the unit cell, the quasi-BIC with high $Q$ factor and tight field confinement are observed. Boosted by the quasi-BIC, the metasurfaces produces a high THG efficiency of $3.9\times10^{-4}$ W$^{-2}$ at the pump wavelength of 1556 nm and outperforms the reported silicon nanodisk resonators. We also show that the FHG can be enhanced with efficiency of up to $4.8\times10^{-10}$ W$^{-4}$ at a moderate pump intensity. Most importantly, the quasi-BIC resonance and the resultant nonlinear harmonics are independent on pump polarization, showing great prospects in designing metasurface and metadevices for nonlinear applications.

\begin{acknowledgments}	
	
This work is supported by the National Natural Science Foundation of China (Grants No. 11947065 and No. 61901164), the Natural Science Foundation of Jiangxi Province (Grant No. 20202BAB211007), the Interdisciplinary Innovation Fund of Nanchang University (Grant No. 2019-9166-27060003), and the China Scholarship Council (Grant No. 202008420045). The authors would like to thank T. Ning, L. Huang, and T. Guo for beneficial discussions on the nonlinear numerical simulations.

\end{acknowledgments}


%

\end{document}